\documentclass[twocolumn,secnumarabic,amssymb, nobibnotes, aps, prd]{revtex4-2}

 \usepackage{amsmath,amsthm,amssymb}
\usepackage[ruled,vlined]{algorithm2e}
\SetKw{KwBy}{by}
\usepackage{graphicx}
\usepackage{tikz}
\usepackage{pgfplots}
\usepgfplotslibrary{groupplots}
\usetikzlibrary{calc}
  \usepackage{xcolor}

\DeclareMathAlphabet{\itbf}{OML}{cmm}{b}{it}

\def \PP{\mathbb P}
\newcommand{\ZZ}{\mathbb{Z}}
\newcommand{\EE}{\mathbb{E}}
\newcommand{\RR}{\mathbb{R}}
\newcommand{\eps}{\epsilon}

\usepackage{physics}
\usepackage{ifthen}
\usepackage{tikz}
\usepackage[outline]{contour} 
\usetikzlibrary{calc} 
\usetikzlibrary{angles,quotes} 
\usetikzlibrary{patterns}
\tikzset{>=latex} 
\contourlength{1.2pt}

\colorlet{myred}{red!65!black}
\tikzstyle{ground}=[preaction={fill,top color=black!10,bottom color=black!5,shading angle=20},
                    fill,pattern=north east lines,draw=none,minimum width=0.3,minimum height=0.6]
\tikzstyle{mass}=[line width=0.6,red!30!black,fill=red!40!black!10,rounded corners=1,
                  top color=red!40!black!20,bottom color=red!40!black!10,shading angle=20]
\tikzstyle{rope}=[brown!70!black,line width=1.2,line cap=round] 

\tikzstyle{force}=[->,thick,line cap=round]
\tikzstyle{Fproj}=[force]

\newboolean{showforces}
\setboolean{showforces}{true}

    \usepackage{combelow}
    
\begin{document}

\title{Computing the diffusivity of a particle subject to dry friction with colored noise}%

\author{Josselin Garnier}%
\email[Josselin Garnier: ]{josselin.garnier@polytechnique.edu}
\affiliation{Centre de Math\'ematiques Appliqu\'ees, Ecole Polytechnique, Institut Polytechnique de Paris, 91120 Palaiseau, France}

\author{Laurent Mertz}%
\email[Laurent Mertz: ]{lmertz@cityu.edu.hk}
\affiliation{Department of Mathematics,  City University of Hong Kong, Kowloon, Hong Kong, China}

\date{\today}%

\begin{abstract}
This paper considers the motion of an object subjected to a dry friction and an external random force.
The objective is to characterize the role of the correlation time of the external random force.
We develop efficient stochastic simulation methods for computing the diffusivity (the linear growth rate of the variance of the displacement) and other related quantities of interest when the external random force is white or colored. These methods are based on original representation formulas for the  quantities of interest which make it possible to build unbiased and consistent estimators.
The numerical results obtained with these original methods are in perfect agreement with known closed-form formulas valid in the white noise regime.
In the colored noise regime the numerical results show that the predictions obtained from the white-noise approximation are reasonable for quantities such as the histograms of the stationary velocity but can be wrong for the diffusivity unless the correlation time is extremely small.
\end{abstract}

%


\maketitle

\newpage

\section{Introduction}
The present work is motivated by the study of the motion of an object subjected to a dry friction and an external random force.
The dry friction model in our paper is the standard model to study macroscopic systems involving solid-solid friction \cite{bowden,persson}.

\begin{figure}[h!]
\includegraphics[scale=0.95]{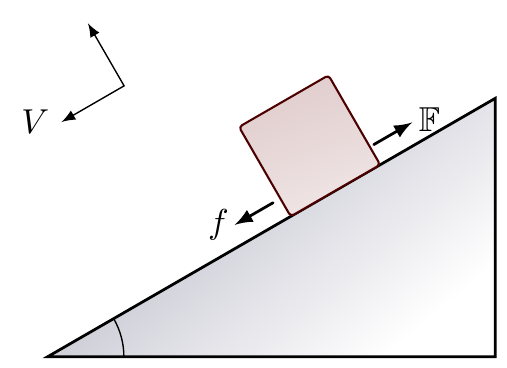}
\caption{Schematic of a solid object drifting downwards on a flat inclined support by overcoming the forces of dry friction~$\mathbb{F}$.}
\label{fig:sliding_object}
\end{figure}  

This dry friction model is rather well understood when the external random force is a white noise \cite{DG05,T10,T12}. 
The probability distribution of functionals of the velocity or the position can 
then be studied in detail \cite{chen14a,chen14b}.
Different generalizations have been considered, such as the motion of a particle bound to a spring being pulled at a definite speed, moving on a surface with dry friction in a noisy environment \cite{feghhi22}.  
Moreover, emerging applications are found for biological systems. The effects of diffusion on the dynamics of a single focal adhesion at the leading edge of a crawling cell are investiaged in  \cite{PhysRevE.102.022134} by considering a simplified model of sliding friction. To understand the stick-slip dynamics of migrating cells on viscoelastic substrates, a theoretical model of the leading edge dynamics of crawling cells is introduced in \cite{PhysRevE.100.012409}.

In our paper, we address the role of the correlation time of the external force when it is a colored noise. 
No explicit formula is available and therefore the analysis goes through numerical simulations.  
Nonetheless, it should be pointed out that an approximate expression of the stationary probability density function of the velocity has been proposed in \cite{GJ17}. 
Our goal is to present appropriate stochastic algorithms to estimate the quantities of interest and to discuss the relationships between the quantities of interest such as the displacement mobility and diffusivity and the input parameters such as the noise strength and correlation time.

We consider the one-dimensional displacement $U$ of an object (with unit mass) lying on a motionless surface. The velocity is denoted by $V$ and thus $V = \dot U$,
where the dot stands for time derivative throughout the paper.
As shown schematically in Figure \ref{fig:sliding_object}, Newton's law of motion implies $\dot V + \mathbb{F} = f$
where $\mathbb{F}$ is the force of dry friction and $f$ represents all the other
external and internal forces, including random perturbations. The force $\mathbb{F}$ cannot be
expressed in terms of a standard function but as follows
$$
 \mathbb{F} = 
 \begin{cases}
 f, \: & \mbox{when} \: V=0 \mbox{ and } |f| \leq \Delta,\\
 \sigma \Delta, \: & \mbox{when} \: V\neq0 \mbox{ or } (V = 0 \mbox{ and } |f| > \Delta),
 \end{cases}
$$
where $\sigma = \textup{sign}(V)$ when $V\neq0$, otherwise $\sigma = \textup{sign}(f)$.
The coefficient $\Delta>0$ is the coefficient of dry friction.
The random perturbation induces a random displacement, thus we can define the diffusivity of the displacement $U$ as
\begin{equation}
\label{diffu_continuous}
D = \lim \limits_{t \to +\infty} D^t \: \mbox{ where } \: D^t = \frac{{\rm var}(U(t))}{t}.
\end{equation}
Such a friction model has been discussed by de Gennes \cite{DG05}. When $f = \sqrt{\Gamma} \dot W$, where $\dot W$ is a white noise (i.e., the time derivative of Brownian motion $W$) and  $\Gamma>0$ is the noise strength, he formally proposed an expansion of the transition probability density of $V$ in terms of eigenmodes related to a one-dimensional Schr\"odinger equation where the potential contains an attractive delta function. As a consequence, he obtained an approximate formula for the correlation function of the velocity. From this formula he suggested that the diffusivity scales as $D \sim \Gamma^3/\Delta^4$. A similar scaling was already proposed in a much earlier work by Caughey and Dienes \cite{CD61}. It is more sensitive to the noise power in contrast to the case where dry friction is replaced by viscous friction, that is $\Delta=0$ and $f = -\tau_L^{-1}V + \sqrt{\Gamma} \dot W$ with $\tau_L>0$ is a relaxation~time. Indeed in this case $D = \tau_L^2 \Gamma$.

Touchette \cite{T10,T12} extended de Gennes' work and obtained without any approximation both the time-dependent transition probability density function and the correlation function of the velocity by solving the associated time-dependent Fokker Planck equation. Touchette's results are exact when $f = \sqrt{\Gamma} \dot W$ or based on series representation when $f = -\tau_L^{-1}V + \sqrt{\Gamma} \dot W$, but they do not cover the case of colored noise.
These results, however,  will be important to us because the stochastic simulation methods that we propose in our paper can be applied in particular to Touchette's configurations and the results deduced from our simulations can, therefore, be tested against exact formulas for these configurations. Our simulation methods, however, can be applied to more general configurations and will unravel behaviors not covered by the previously known formulas.

Goohpattader et al. \cite{GMC09} have experimentally investigated physical friction problems that can be modeled using the aforementioned framework. They considered a forcing of the form $f = -\tau_L^{-1} V + \bar{\gamma} + \sqrt{\Gamma} \dot W$ 
where  $\bar{\gamma}$ is a constant related to gravity and the inclination of the surface on which the system is installed.
They also proposed numerical simulations. They observed experimentally and by simulation that the variance of the object displacement grows linearly with time
and they also observed scaling laws for the diffusivity that we will challenge in our paper.

Recently, some of the authors of the present paper
have considered the case where 
$f=\mathfrak{b}(V) + \sqrt{\Gamma} X$,
$\mathfrak{b}(.)$ is a general function with appropriate conditions, and $X$ is a pure jump noise (i.e., a piecewise constant random process). In \cite{GLM23} 
they proposed a piecewise deterministic Markov process (PDMP) to model the pair $(X,V)$. 
This framework makes it possible to use the theory and simulation methods of PDMPs \cite{davis84,saporta}. They derived the Kolmogorov equations for the pair $(X,V)$. When $\mathfrak{b}(.)$ is an odd function, they showed ergodicity and provided a representation formula of the stationary state in terms of a portion of the trajectory called short excursion.
Essentially, a short excursion contains only one dynamic phase, a time interval on which $V \neq 0 \mbox{ or } |f| > \Delta$, and only one static phase, a time interval on which $V=0 \mbox{ and } |f| \leq \Delta$. 

We develop our present article on the basis of the PDMP framework mentioned above and introduce new stopping times which identify independent components in the dynamics. These components are different from the short excursions. We call them \textit{long excursions}. Having identified this type of trajectory portion we can express the diffusivity (or any related quantity) as an expectation of a functional of a long excursion 
and we can, therefore, estimate these quantities by sampling long excursions directly instead of sampling long-time period integrals on the original PDMP. 
We finally extend the notion of long excursion together with the corresponding sampling method to the limiting system case when the time step of the PDMP  goes to zero. The latter is formulated using a differential inclusion \cite{B77,MR3308895} forced by a colored noise.
The estimators based on our stochastic simulation methods are unbiased contrarily to the standard estimation methods that consist in taking long but fixed-length trajectories.
They are consistent and asymptotically normal.
Their accuracies are sufficient to be used to discuss quantitive relations between the diffusivity and the noise strength and correlation time.
In particular, they show that the predictions for the values of the diffusivity obtained from the white-noise approximation can be wrong when the correlation time of the noise is moderately small.

This paper is organized as follows: 
Section \ref{sec:dim} proposes a dimensional analysis of the system in order to identify its effective parameters. Section \ref{sec:pdmp} describes the PDMP framework modeling the friction problem and defines the original notion of long excursion. Section \ref{sec:formula} presents our new characterization of the displacement diffusivity using long excursions. The resulting algorithm and an ad hoc Monte Carlo estimator are proposed in Section \ref{sec:mc}. 
In Section \ref{sec:limit}, the notion of long excursion and the resulting numerical approach are extended from the PDMP case to the limiting differential inclusion case.
Numerical simulations for the relation between the noise strength and correlation time and the diffusivity are studied in Section \ref{sec:num}.  Finally, we conclude in Section \ref{sec:conclusion}.

\section{Effective parameters and non-dimensional system}
\label{sec:dim}%
The driving noise with a correlation time $\tau>0$ is represented by $X$ and the resulting velocity $V$ 
satisfies, using the notation $f = \mathfrak{b}(V) + \sqrt{\Gamma} X$ with $\mathfrak{b}(v)$ a Lipschitz continuous function, 
\begin{equation}
\label{puredryfriction0}
\begin{cases}
\dot V = f - \sigma \Delta, \: & \mbox{ when } \: V \neq 0 \: \mbox{ or } \: |f|>\Delta \mbox{ (dynamic phase)}, \\
\dot V = 0, \: & \mbox{ when } \: V = 0 \: \mbox{ and } \: |f| \leq \Delta \mbox{ (static phase)},
\end{cases}
\end{equation}
where we have denoted 
$\sigma = \textup{sign}(V)$ when $V\neq0$, otherwise $\sigma = \textup{sign}(f)$. 
Equation~(\ref{puredryfriction0}) can equivalently be written in the form of a multivalued stochastic differential equation (MSDE):
\begin{equation}
\label{puredryfriction}
\dot{V}+\partial \varphi(V) \ni \mathfrak{b}(V) +\sqrt{\Gamma} X.
\end{equation}
Here $\varphi(v)=\Delta |v|$ and its subdifferential $\partial \varphi$ is the set-valued map given by $\partial \varphi(0) = [-\Delta,\Delta]$ (interval) 
and $\partial \varphi(V) = \{ {\rm sign}(V) \Delta\}$ (singleton) when $V \neq 0$.
The MSDE is a concise and rigorous way to formulate the transition between static and dynamic phases.
 A gentle introduction to MSDEs can be found in Chapter 4 of \cite{MR3308895}.

Below we derive the effective parameters and the corresponding non-dimensional system. 
We remark that the physical parameters  $\Delta$  and $\Gamma$ are expressed in $m s^{-2}$ and  in $m^2 s^{-3}$, respectively.
We can then introduce the reference time and space units $\tau_0= \Gamma \Delta^{-2}$ (in $s$) and $u_0=\Gamma^2 \Delta^{-3}$ (in $m$). We deduce the non-dimensional variables 
\begin{equation}
t' = t /\tau_0 
, \quad 
V' (t') = V(t' \tau_0)  \tau_0/u_0  
, \quad 
X' (t')= X (t'\tau_0) \tau_0^{1/2} . 
\end{equation}

When $\mathfrak{b}(v) = - v/\tau_L + \bar \gamma$, 
we can recast Equation \eqref{puredryfriction} into the non-dimensional form 
\begin{equation}
\dot V' + \partial |V'| \ni - V'/ \tau_L' + \bar{\gamma}' + X'  ,
\end{equation}
where $\tau_L' = \tau_L /\tau_0$, $\bar{\gamma}'=\bar{\gamma} /(u_0/\tau_0^2) =\bar{\gamma}  / \Delta$, and 
the dot stands for the derivative with respect to $t'$.
Moreover, the effective noise correlation time from the non-dimensional dynamics is
$\tau' = \tau/\tau_0$.
We will discuss the impact of $\tau'$ on statistics of the system.
 
\section{The PDMP system}
\label{sec:pdmp}%
In this section we present the system that describes the motion driven by a dry friction and an external random, stepwise constant force.

\subsection{Description of the pure jump noise}
\label{subsec:markovjump}%
We first define the driving colored noise $X$ as a Markov jump process.

Let $\delta>0$ be a grid step (for the noise).
The process $X$ takes values in the finite
 state space $S^\delta = \delta \ZZ \cap [ - L_X^\delta, L_X^\delta]$, with $L_X^\delta \uparrow +\infty$ 
 as $\delta \downarrow 0$. 
Thus, $S^\delta $  is a finite set of equally $\delta$-spaced points denoted by $\{x_{-N},\ldots,x_N\}$, where $N=[L_X^\delta \delta^{-1}]$. 
We also introduce the non-dimensional spacing $ \delta' = \delta \tau_0^{1/2}= \delta \Delta^{-1} \sqrt{\Gamma}$. 

The process $X$ is stepwise constant over time intervals whose durations are independent and identically distributed with the exponential distribution with parameter  $\Lambda = 2 \tau^{-2} \delta^{-2}$. At the jump times the process randomly jumps to one of its nearest neighbors.
If it is at position $x$, then the process jumps to the right neighbor $x+\delta$ with probability $\alpha_x = \frac{1}{2} \big(1 -\frac{ \tau \delta x}{2}\big)$ and it jumps to the left neighbor $x-\delta $  with probability $1- \alpha_x$
 (except when it is at the boundaries of its state space where it deterministically jumps to its unique nearest neighbor).
The stochastic simulation method to generate trajectories of $X$ is described in Appendix~\ref{app:genX}.

The process $X$ can be seen as a discretization of an Ornstein-Uhlenbeck (OU) process with correlation time $\tau>0$.
In \cite{GLM23} it is proved that the process $X$ converges in distribution  to  $X^\star$ as $\delta \to 0$,  where $ X^\star$ is an OU process, that is solution of the stochastic differential equation
\begin{equation}
\label{noiselimit}
\tau \dot X^\star = -X^\star + \sqrt{2} \dot W,
\end{equation}
with $\dot W$ a white noise. 
The OU process $X^\star$ is a stationary zero-mean Gaussian process with correlation function $\EE[X^\star (0)X^\star (t)]=(1/\tau) \exp(-|t|/\tau)$.
From the dimensional analysis of Section \ref{sec:dim} and the expression of the non-dimensional spacing $ \delta'$, we can actually approximate the distribution of $X$ by the distribution of $X^\star$ when $\delta'$ is much smaller than one.
This means that $X$ is indeed a discretization of the OU process $X^\star$ with correlation time $\tau$.
Additionally, when  $\tau'$ is much smaller than one, then $X^\star$ behaves like the white noise $\sqrt{2}\dot W$.

\subsection{Description of the PDMP}
We now define  the PDMP modeling dry friction driven by the noise $X$. 
The PDMP is the process $Z=(X,Y,V)$. 
The coordinate $X$ is the jump process modeling the driving force described above. 
The coordinate $V$ is the continuous process defined by \eqref{puredryfriction0} or \eqref{puredryfriction}.
The coordinate $Y$ is the jump process determined by 
$Y=\Theta(X,V)$, with 
\begin{equation}
 \Theta(x,v) = 
\left\{
\begin{array}{ll}
1 & \mbox{ if }v>0 \mbox{ or if } v=0,  \, \sqrt{\Gamma} x> -\mathfrak{b}(0)+\Delta,\\ -1& \mbox{ if }v<0 \mbox{ or if } v=0, \, \sqrt{\Gamma} x< -\mathfrak{b}(0)-\Delta,\\ 0 &\mbox{ if }v=0, \, \sqrt{\Gamma} x \in [-\mathfrak{b}(0)-\Delta,-\mathfrak{b}(0)+\Delta].
\end{array}
\right.
\end{equation}
The marker $Y$ indicates whether the process is in a dynamic phase ($|Y|=1$) or in a static phase ($Y=0$).
The introduction of the marker $Y$ makes it possible to adopt the formalism of PDMPs, with smooth flows for the continuous process $V$ and jumps of the mode $(X,Y)$ that occur at random times when $X$ jumps and when the dynamics for $V$ changes from the static to the dynamic phases.
 We give details on the definition of the PDMP $Z$ in Appendix~\ref{app:PDMP}.
This formalism allows to use the theory and simulation methods developed for PDMPs described in \cite{davis84,saporta} and it will allow us to introduce new representation formulas for quantities of interest using strong Markov property.

It is proved in \cite{GLM23} that the random process  $(X,V)$ converges in distribution 
 to the Markov process  $(X^\star,V^\star)$ which is solution of \eqref{noiselimit}-\eqref{puredryfriction}.
So we can consider the process $(X,V)$ as a discretization of the process $(X^\star,V^\star)$.

 \subsection{Definition of long excursions}
 \label{sec:deflongexcursion}
\begin{figure}[h!]
\centering
\includegraphics[scale=1]{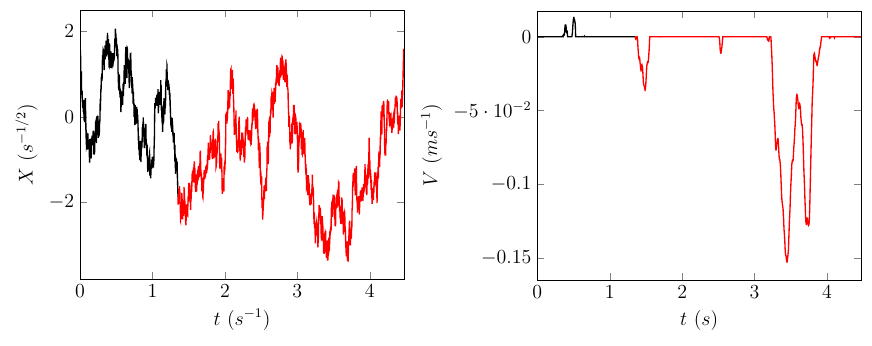}
\includegraphics[scale=1]{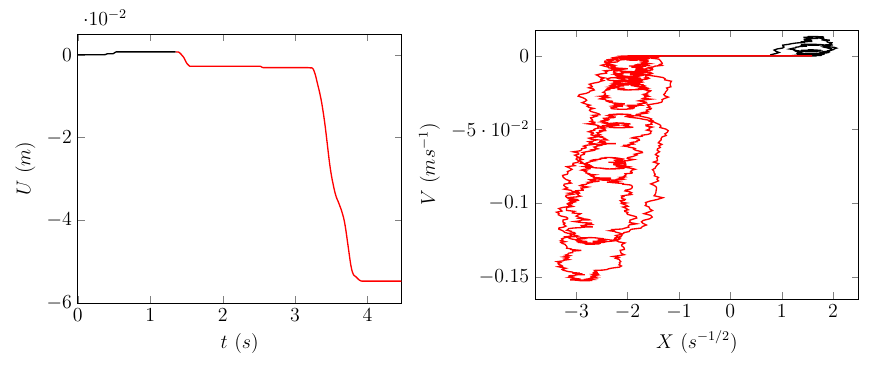}
\caption{Numerical simulation of a long excursion of $(X,V)$ enclosed by the random time interval $[0,t_1]$. The first HLE in black is followed by the second HLE in red. 
Top left: noise $X$ versus time $t$. Top right: velocity $V$ versus time $t$. Bottom left: displacement $U$ versus time $t$.  Bottom right:  velocity $V$ versus noise $X$. 
Here $\mathfrak{b}(v) = - v/\tau_L + \bar \gamma$, $\tau = 0.5 \, s$,  $\tau_L = 0.067 \, s$, $\Delta = 3.84 \, m s^{-2}$, $\Gamma = 5 \, m^2 s^{-3}$, and $\bar \gamma = 0.342 \, ms^{-2}$. This sample is produced by Algorithm~\ref{algo:1} with $\delta = 0.125 \, s^{-1/2}$.} 
\label{fig1}
\end{figure}

A long excursion is composed of two parts which we call half-long excursions (HLE). 
We define the two integers $k_-$ and $k_+$ by $ \sqrt{\Gamma}x_{k_+} \leq -\mathfrak{b}(0) + \Delta <  \sqrt{\Gamma}x_{k_++1}$ and $ \sqrt{\Gamma}x_{k_--1} <  -\mathfrak{b}(0) - \Delta \leq  \sqrt{\Gamma} x_{k_-}$.
The two integers $k_-$ and $k_+$ play important roles because a transition from a static phase to a dynamic phase occurs when $Z$ jumps from $(x_{k_+},0,0)$ to $(x_{k_++1},1,0)$ or from  $(x_{k_-},0,0)$ to $(x_{k_--1},-1,0)$. 
We can define the first HLE originating from $(x_{k_++1},1,0)$ as a portion of trajectory of the process $Z$ starting from $(x_{k_++1},1,0)$ at time $0$ and ending in $(x_{k_--1},-1,0)$ at time 
$t_{\frac{1}{2}} = \inf \{ t \geq 0, \: V(t) = 0 \: \mbox{ and } \: X(t) = x_{k_--1} \}$. The second HLE starts from $(x_{k_--1},-1,0)$ at time $t_{\frac{1}{2}}$ and ends in $(x_{k_++1},1,0)$ at the time 
$t_1 = \inf \{ t \geq t_{\frac{1}{2}}, \: V(t) = 0 \: \mbox{ and } \: X(t) = x_{k_++1} \}$.
We use the notation $\pm$-HLE for a half-long excursion originated from $(x_{k_\pm \pm 1},\pm 1,0)$ (see Figure~\ref{fig1}).
In general, a long excursion is defined as the concatenation of $\pm$-HLE followed by a $\mp$-HLE. It is worth noting that it is possible that such an HLE evolves only in a dynamic phase. Long excursions are building blocks for the forthcoming representation formulas for quantities of interest such as the diffusivity.

\section{Mobility and diffusivity}
\label{sec:formula}
In this section, we propose original representation formulas for the displacement mobility and diffusivity in terms of a long excursion.
These formulas will then be used to build efficient estimators of the diffusivity in the next section.
We consider the displacement $U(t)$. 
It satisfies the two following properties:\\
{\bf 1)} $U(t)/t$  converges in probability as $t \to +\infty$ to  
\begin{equation}
M_0 = \frac{\EE_{s_+}[U(t_1)]}{\EE_{s_+}[t_1]},
\label{eq:defM0}
\end{equation}
where $\EE_{s_+}$ stands for the expectation with respect to the distribution of the PDMP starting from $s_+ = (x_{k_++1},1,0)$.\\
{\bf 2)} ${\sqrt{t}} (U(t)/t -M_0)$ converges in distribution as $t \to +\infty$ to a Gaussian variable with mean zero and variance
\begin{equation}
\label{eq:defD}
D = \frac{{\rm Var}_{s_+}   \big(U(t_1)\big)}{\EE_{s_+}[t_1]}  .
\end{equation}
We will show in the following sections that the two representation formulas (\ref{eq:defM0}) and (\ref{eq:defD}) make it possible to build unbiased and consistent Monte Carlo estimators.
The remainder of this section is devoted to the proof of (\ref{eq:defM0}) and (\ref{eq:defD}), 
which is based on standard limit theorems (law of large numbers and central limit theorem) and strong Markov property.

{\it Proof of (\ref{eq:defM0}) and (\ref{eq:defD}).}
We introduce $s_- = (x_{k_--1},-1,0)$, $t_0=0$, and for $j\geq 0$: $t_{j+1} =\inf \{ t \geq t_{j+1/2}, (X(t),Y(t),V(t)) = s_+\}$, $t_{j+1/2}=\inf \{ t \geq t_j , (X(t),Y(t),V(t))=s_- \}$,
\begin{align*}
J_t = \inf \{ j \geq 1, \, {t}_j \geq t \},
\qquad
j_t = \left\lfloor \frac{t}{\EE_{s_+}[t_1]}\right\rfloor,
\end{align*}
where $\lfloor\cdot\rfloor$ stands for the integer part.
The random variables 
$$
{\mathcal X}_j = \int_{{t}_j}^{{t}_{j+1}} V(s) ds 
$$
are independent and identically distributed as ${\mathcal X}_0=U(t_1)$ under $\EE_{s_+}$ by the strong Markov property. If $\mathfrak{b}$ is an odd function, then $\EE_{s_+}[{\mathcal X}_0]=0$ (this can be shown by a symmetry argument, because $(X_t,V_t)_{t\geq 0}$ has then the same distribution as $(-X_t,-V_t)_{t \geq 0}$),
but in general it is not zero.


We have 
$$
\frac{1}{t}U(t)
=\frac{ {j_t}}{ t} \Big[ \frac{1}{ {j_t}}
\sum_{j=0}^{j_t-1} {\mathcal X}_{j} + R_t \Big],
$$
with
$$
R_t =  \frac{1}{ {j_t}}\int_{{t}_{j_t}}^t V(s) ds .
$$
We show in Appendix \ref{app:A} that $\sqrt{t} R_t$, hence  $R_t$, converges in probability to zero as $t\to +\infty$.
Moreover, $j_t\to \infty$ as $t\to+\infty$ so we obtain from the Law of Large Numbers  that $  {j_t}^{-1}
\sum_{j=0}^{j_t-1} {\mathcal X}_{j} $ converges in probability to $\EE_{s_+}[{\mathcal X}_0]$.
We also observe that $ {j_t} /  {t} \to 1/ { \EE_{s_+}[t_1]}$. Therefore, we obtain:
\begin{equation}
\frac{1}{ {t}} U(t)
\stackrel{proba.}{\longrightarrow}  M_0 =\frac{ \EE_{s_+}[{\mathcal X}_0] }{\EE_{s_+}[t_1] },
\end{equation}
which gives (\ref{eq:defM0}).
In order to show (\ref{eq:defD}), we write
$$
 \sqrt{t} \Big( \frac{U(t)}{t} - M_0 \Big)
=\frac{\sqrt{j_t}}{\sqrt{t}} \Big[ \frac{1}{\sqrt{j_t}}
\sum_{j=0}^{j_t-1} \tilde{\mathcal X}_{j} + \tilde{R}_t \Big]+ \tilde{r}_t,
$$
with $\tilde{\mathcal X}_{j}= {\mathcal X}_{j} -\EE_{s_+}[{\mathcal X}_0 ]$,   
\begin{align*}
\tilde{R}_t &=  \sqrt{t}R_t=\frac{1}{\sqrt{j_t}}\int_{{t}_{j_t}}^t V(s) ds , \\
\tilde{r}_t &= \frac{j_t}{ \sqrt{t}  } \EE_{s_+}[{\mathcal X}_0 ] -\sqrt{t} M_0 .
\end{align*}
We show in Appendix \ref{app:A} that $\tilde{R}_t$ converges in probability to zero as $t\to +\infty$.
The quantity $\tilde{r}_t =  \EE_{s_+}[{\mathcal X}_0 ] ( \lfloor t/ \EE_{s_+}[t_1 ] \rfloor- t/ \EE_{s_+}[t_1 ]  )/\sqrt{t}$ is such that $|\tilde{r}_t | \leq | \EE_{s_+}[{\mathcal X}_0 ]|/\sqrt{t}$ so it also converges to zero as $t\to +\infty$.
Since $\tilde{\mathcal X}_{j}$ are independent and identically distributed with mean zero, we obtain from
the Central Limit Theorem  that ${j_t}^{-1/2}
\sum_{j=0}^{j_t-1} \tilde{\mathcal X}_{j} $ converges in distribution to a zero-mean Gaussian variable with variance ${\rm Var}_{s_+}({\mathcal X}_0)$.
We also observe that $\sqrt{j_t} / \sqrt{t} \to 1/\sqrt{ \EE_{s_+}[t_1]}$. By Slutsky's theorem we obtain:
\begin{equation}
 \sqrt{t} \Big( \frac{U(t)}{t} - M_0 \Big)
\stackrel{dist.}{\longrightarrow} {\mathcal N} (0 , D), \quad D= \frac{ {\rm Var}_{s_+}({\mathcal X}_0)}{\EE_{s_+}[t_1]},
\end{equation}
which completes the proof of the desired result.
\qed 

\section{Monte Carlo estimation of the diffusivity}
\label{sec:mc}
\subsection{Monte Carlo estimator based on long excursions}
Consider a long excursion as defined in Subsection~\ref{sec:deflongexcursion}. It is composed of two HLEs. 
We  can now introduce an original Monte Carlo method for the estimation of $D$. Let $(U^{(k)}_{\textup{le}},t^{(k)}_{\textup{le}})$, $k=1,\ldots,N$, be $N$ independent and identically distributed (i.i.d.) pairs of displacement $U(t_1)$ and duration $t_1$ both resulting from a long excursion. We introduce a Monte Carlo estimator based on long excursions as follows:
\begin{equation}
\label{le_estimator}
\hat{D}_N = \frac{\sum_{k=1}^N  \left ( U^{(k)}_{\textup{le}} \right )^2 - \frac{1}{N} \left (\sum_{k=1}^N U^{(k)}_{\textup{le}} \right )^2}{\sum_{k=1}^N t^{(k)}_{\textup{le}}}.
\end{equation}
The sample $\{ (U^{(k)}_{\textup{le}},t^{(k)}_{\textup{le}}) \}_{k=1}^N$ is produced by using Algorithm~\ref{algo:1}.
The estimator $\hat{D}_N$ is consistent by the Law of Large Numbers.
Beyond the estimator $\hat{D}_N$, it is possible to build from the sample $\{ (U^{(k)}_{\textup{le}},t^{(k)}_{\textup{le}}) \}_{k=1}^N$
a confidence interval with prescribed asymptotic confidence level $\alpha$ (see Appendix \ref{app:B}).

\begin{algorithm}[H]
\label{algo:1}
\SetAlgoLined
\KwResult{Simulation of  $\left \{(X,Y,V)_{T_j} 
\mbox{ where } \: j \geq 0 \: \mbox{ and }
\: T_j \leq t_{\frac{1}{2}}
\right \}$. 
}
 $T  = 0, 
 \: X  = x_{k_++1}, 
 \: Y  = 1,
 \: V = 0, 
 \: U = 0,
 \: A =$ TRUE\;
  
\While{$A$}{
$\delta T =$ interjump$(X,Y,V)$\;
$U = U +$displacement$(X,Y,V,T,T+\delta T)$\;
$(X,Y,V) =$ jump$(X,Y;\textup{flow}(X,Y,V; \delta T))$\;
$T= T + \delta T$\;
$A = (X \neq x_{k_- -1})$ or $(Y \neq -1)$ or $(V \neq 0)$\;
}

\caption{PDMP simulation for the first HLE from $(x_{{k_+}+1},1,0)$ to $(x_{{k_-}-1},-1,0)$.}
\end{algorithm}
To simulate the other HLE, we can swap $x_{{k_+}+1}$ with $x_{{k_-}-1}$, $(Y = 1)$ with $(Y=-1)$, and vice versa in Algorithm \ref{algo:1}.
The functions interjump$(X,Y,V)$, displacement$(X,Y,V)$, \textup{flow}$(X,Y,V,\delta T)$ (which is used in displacement$(X,Y,V)$) and jump$(X,Y ; V)$ are described in Appendix \ref{app:C}.

\subsection{Brute force Monte Carlo estimator}
For comparison, we also consider the brute force Monte Carlo estimator for $D$, that is
\begin{equation}
\hat D_{N'}^t = \frac{1}{t} \bigg[ \frac{1}{N'} \sum_{k=1}^{N'} U^{(k)}(t)^2 - \Big( \frac{1}{N'} \sum_{k=1}^{N'} U^{(k)}(t) \Big)^2 \bigg], 
\end{equation}
here the sample $\{ U^{(k)}(t) \}_{k=1}^{N'}$ is composed of ${N'}$ i.i.d. realizations of the displacement at time $t$ and 
is produced by using Algorithm~\ref{alg2}.
Note that $\hat{D}^t_{N'}$ is actually a consistent estimator of $D^t$. This means that $t$ should be chosen large enough so that the bias (the difference between $D^t$ and $D$) is negligible. We discuss this point in detail in Subsection \ref{subsec:eff}.

\begin{algorithm}[H]
\label{alg2}
\SetAlgoLined
\KwResult{Simulation of  $\left \{(X,Y,V)_{T_j} 
\mbox{ where } \: j \geq 0 \: \mbox{ and }
\: T_j \leq t
\right \}$.
}
 $T  = 0, 
 \: X  =  x_{k_++1}, 
 \: Y  = 0,
 \: V = 0, 
 \: U = 0$\;
  
\While{$ (T<t)$}{
$\delta T =$ interjump$(X,Y,V)$\;
$U = U +$ interjump$(X,Y,V)$\;
$(X,Y,V) = $jump$(X,Y;$flow$(X,Y,V; \delta T))$\;
$T= T + \delta T$\;
{\bf if} ($T\geq t$) {\bf then} $U = U +$ displacement$(X,Y,V,T-\delta T,t)$\;
}
\caption{PDMP simulation on $[0,t]$.}
\end{algorithm}


\subsection{Asymptotic efficiencies of the estimators}
\label{subsec:eff}
In this section we show that the mean square error of the estimator $\hat{D}_N$ based on long excursions is much smaller than the one of the brute force Monte Carlo estimator $\hat{D}_N^t$ even when tuning the parameter $t$ optimally.

From the delta method (described in Appendix~\ref{app:B}),
the mean square error of the estimator $\hat D_N$ satisfies 
\begin{equation}
\mathbb{E}_{s_+} \Big[ \big(\hat D_N - D \big)^2 \Big]\sim \frac{\sigma^2}{N}  ,
\end{equation}
as $N\to +\infty$,
where the variance $\sigma^2 = \nabla \Psi({\itbf S})^T {\bf C} \nabla \Psi({\itbf S})$ involves 
${\itbf S} = \EE_{s_+}[{\itbf X}]$, ${\bf C}=(C_{jl})_{j,l=1}^3$, $C_{jl} = \EE_{s_+}[ X_j X_l]-\EE_{s_+}[ X_j ]\EE_{s_+}[X_l]$, 
$\Psi({\itbf x})=\frac{x_2-x_1^2}{x_3}$
with ${\itbf X}=(X_j)_{j=1}^3$, $X_1=U_{\textup{le}} = U(t_1)$, $X_2= {U_{\textup{le}}}^2 = U(t_1)^2$, 
$X_3=t_{\textup{le}} = t_1$.

The mean square error of the estimator $\hat D_{N'}^t$ satisfies
\begin{equation}
\mathbb{E}_{s_+}  \Big[ \big( \hat D_{N'}^t - D  \big)^2 \Big] \sim \frac{(\sigma^t)^2}{N'} + (D^t-D)^2 ,
\end{equation}
as $ N' \to \infty$,
where  $(\sigma^t)^2 = \nabla \Phi({ \itbf R})^T {\bf \Gamma} \nabla \Phi({ \itbf R})$ 
involves 
${\itbf R} = \EE_{s_+}[{\itbf Y}]$, ${\bf \Gamma}=(\Gamma_{jl})_{j,l=1}^2$, $\Gamma_{jl} = \EE_{s_+}[ Y_j Y_l]-\EE_{s_+}[ Y_j ]\EE_{s_+}[Y_l]$, 
$\Phi({\itbf y})=y_2-y_1^2$
with ${\itbf Y}=(Y_j)_{j=1}^2$, 
$Y_1 = U(t)/\sqrt{t}$, 
$Y_2 = U(t)^2/t$.
Note that the mean square error is the sum of a variance term and a squared bias term. The latter turns out to have a dramatic effect.

Denoting $\overline{t_1}= \mathbb{E}_{s_+}(t_1)$, it takes 
$\sum_{k=1}^N t_{\textup{le}}^{(k)}  \approx N \overline{t_1}$ computational time units to produce the sample $\{ (U_{\textup{le}}^{(k)},t_{\textup{le}}^{(k)}) \}_{k=1}^N$ and $N't$ computational time units to produce the sample  $\{ (U^{(k)}(t) \}_{k=1}^{N'}$.
Therefore, when $t = \alpha\overline{t_1}$, we consider the relation $N' \alpha = N$ in order to compare $\hat D_N$ and $\hat D_{N'}^t $(which becomes $\hat{D}_{N/\alpha}^{\alpha \overline{t_1}}$) with identical computational cost. With $t = \alpha\overline{t_1}$, the mean square error of the estimator $\hat D_{N/\alpha}^{\alpha \overline{t_1}}$ satisfies
$$
\mathbb{E}_{s_+}  \Big[ \big( \hat D_{N/\alpha}^{\alpha\overline{t_1}} - D \big)^2 \Big]
\sim 
\frac{\alpha (\sigma^{\alpha \overline{t_1}})^2}{N} + (D^{\alpha \overline{t_1}}-D)^2 ,
$$
as  $N \to \infty$.
We want to compare the mean square errors of the estimators $\hat{D}_N$  and $\hat{D}_N^t$. First, we need to tune the parameter $t$ to get the minimal error.

We first consider the case when $M_0=0$.
When $\alpha$ becomes large, $(\sigma^{\alpha \overline{t_1}})^2$ converges to $2D^2$.
When $\alpha$ becomes large, we have $D^{\alpha \overline{t_1}}=D+O(\alpha^{-1})$. Indeed, 
$$
D^t=\frac{2}{t}\int_0^{t}  \int_0^{t-s} {\rm Cov}_{s_+} (V(s),V(s+s')) {\rm d}s' {\rm d}s,
$$
${\rm Cov}_{s_+} (V(s),V(s+s'))$ converges exponentially as $s\to +\infty$ to an integrable function $\phi(s')$ which is the stationary covariance function of $V$ (see Figure~\ref{fig:4}) and 
$D= \lim_{t \to +\infty} D^t = 2\int_0^\infty\phi(s') {\rm d} s'  $ so that 
\begin{align*}
& \frac{t}{2}(D^t -D)\\
&
=  \int_0^t \int_0^\infty \big[ {\rm Cov}_{s_+} (V(s),V(s+s')) {\bf 1}_{s'<t-s}   - \phi(s')  \big] {\rm d}s' {\rm d}s  \\
&\stackrel{t \to +\infty}{\longrightarrow} 
 \int_0^\infty  \int_0^\infty \big[ {\rm Cov}_{s_+} (V(s),V(s+s'))    - \phi(s')  \big] {\rm d}s' {\rm d}s  \\
 &\qquad -
\int_0^\infty s' \phi(s') {\rm d} s' .
\end{align*}
For $t=\alpha \overline{t_1}$ we define $\alpha_N^\star$ the minimizer of the function $\alpha \mapsto  N^{-1} \alpha (\sigma^{\alpha \overline{t_1}})^2 +(D^{\alpha \overline{t_1}}-D)^2$.
By the two previous observations about the asymptotic behaviors of $\sigma^{\alpha \overline{t_1}}$ and $D^{\alpha \overline{t_1}}-D$, we find that $\alpha_N^\star$ is of the order of  $\alpha_N^\star \sim N^{1/3}$ so that the minimal mean square error of $ \hat D_{N/\alpha}^{\alpha \overline{t_1}}$ obtained with $\alpha_N^\star$ is of order $N^{-2/3}$. That means that, even when tuning the brute force Monte Carlo with the optimal $t$, 
its means square error is larger than the mean square error of $ \hat D_{N}$ which is of order $N^{-1}$ without any tuning. This shows that the estimator $ \hat D_{N}$ is clearly preferable in  the regime when $N$ is large.

When $M_0 \neq 0$, the situation is even worse for the brute force Monte Carlo estimator, because $(\sigma^t)^2$ becomes equivalent to $M_0^2 D t$ for large $t$, so that the optimal $\alpha_N^\star \sim N^{1/4}$ and  the minimal mean square error of $ \hat D_{N/\alpha}^{\alpha \overline{t_1}}$ obtained with $\alpha_N^\star$ is of order $N^{-1/2}$.

This is the main output of this paper from the methodological point of view: the estimation of the diffusivity (or mobility or any other asymptotic quantity) should be carried out with the Monte Carlo method based on long excursions rather than the Monte Carlo method based on long fixed-time excursions that is traditionally used in the literature.


\section{Limiting differential inclusion}
\label{sec:limit}
In this section, the notion of long excursion and the corresponding sampling approach are extended to the limiting differential inclusion case.

\subsection{Long excursion of the differential inclusion}
From \cite{GLM23}, the PDMP $(X,V)$ converges in distribution as $\delta \to 0$ towards $(X^\star,V^\star)$ the solution of the differential inclusion: 
\begin{equation}
\label{eq:diff_incl}
\dot V^\star + \partial \varphi (V^\star) \ni \mathfrak{b}(V^\star) 
+ \sqrt{\Gamma} X^\star, 
\end{equation}
where $\tau \dot X^\star = - X^\star + \sqrt{2} \dot W$ and $\varphi(v) = \Delta |v|$. The process $X^\star$ is an OU process 
 and its invariant density is a Gaussian distribution with mean zero and variance $\tau^{-1}$.

It is natural to extend the concepts of long excursion to the limiting differential inclusion case. The definitions of half-long and long excursions for $(X^\star,V^\star)$ are similar to those of $(X,V)$ defined in section \ref{sec:deflongexcursion}.
The first HLE for $(X^\star,V^\star)$ starts at time $0$ from $(x_\Delta,0)$ and ends at time 
$
t_{1/2}^\star =  \inf \{ t \geq 0, X^\star(t) = x_{-\Delta} \: \mbox{ and } \: V^\star(t) = 0 \}.
$   
Then, the second HLE for $(X^\star,V^\star)$ starts at time $t_{1/2}^\star$ from $(x_{-\Delta},0)$ and ends at time 
$
t_1^\star =  \inf \{ t \geq t_{1/2}^\star, X^\star(t) = x_{\Delta} \: \mbox{ and } \: V^\star(t) = 0 \}.
$   
Here, we have introduced the points $x_\Delta=(-\mathfrak{b}(0)+\Delta)/\sqrt{\Gamma}$ and  $x_{-\Delta}=(-\mathfrak{b}(0)-\Delta)/\sqrt{\Gamma}$.

The diffusivity $D^\star$ is defined as in (\ref{diffu_continuous}) but with $\dot{U}^\star=V^\star$. It has the representation formula in terms of the long excursion:
\begin{equation}
\label{eq:represDstar}
D^\star = \frac{{\rm Var}_{(x_\Delta,0)}   \big(U^\star(t_1^\star)\big)}{\EE_{(x_\Delta,0)}[t_1^\star]}  .
\end{equation}

\subsection{Monte Carlo estimator}
In this section we define a MC estimator $\hat D_N^\star$  of the diffusivity $D^\star$. This estimator is based on the representation formula (\ref{eq:represDstar}) in  terms of the long excursions of the differential inclusion (in a similar manner to what was done for $\hat D_N$). 

Let $\{ U_{\textup{le}}^{\star,(k)}, t_{\textup{le}}^{\star,(k)}\}_{k=1}^N$ be $N$ i.i.d. pairs of displacement and duration resulting from a long excursion. This sample is produced by Algorithm~\ref{algo_le_diff_inclu}. 

\begin{algorithm}[H]
\label{algo_le_diff_inclu}
\SetAlgoLined
\KwResult{$(X^\star,V^\star)$ on the interval $[0,t_{1/2}^\star]$.}
\caption{Differential inclusion simulation for the first HLE from $(x_\Delta,0)$ to $(x_{-\Delta},0)$.}
$
 \: X^\star  = x_{\Delta}, 
 \: V^\star = 0, 
 \: f^\star = 0,
 \: U^\star = 0,
 \: A =$ TRUE\;
  
\While{$A$}{
$(\hat\Xi,\hat{X})^T \sim \mathcal{N}(X^\star m(h),\Sigma(h))$\;
$(\hat f, \hat V) = (f^\star,V^\star)$\;
$f^\star = \hat V + h (\mathfrak{b}(\hat V) 
 + \sqrt{\Gamma} \hat\Xi)$\; 
$V^\star = \hat V - h \max(-\Delta,\min(\Delta, \hat f h^{-1}))$\;
$X^\star= \hat{X}$; 
$U^\star = U^\star + h \hat V$\;
$A = (|\hat f | > \Delta)$ or $(\hat V \neq 0)$ or $f^\star \leq - \Delta$\;

}
\end{algorithm}
Here, the notation $(\hat\Xi,\hat X)^T \sim \mathcal{N}(x m(h),\Sigma(h))$ means that $(\hat\Xi,\hat X)^T$ is a realization of a two-dimensional Gaussian variable with expectation $xm(h)$ with 
$$
m(h) =  \left ( \frac{\tau}{h} (1 - e^{-{h}/{\tau}} )  , e^{-{h}/{\tau}} \right )^T
$$ 
and with covariance matrix
$$
\Sigma(h) =  \begin{pmatrix} \frac{\tau}{h^2} 
\left ( 2 \frac{h}{\tau} - 3 + 4 e^{-h/\tau} - e^{-2 h/ \tau } \right ) & \frac{1}{h} \left ( 1 - e^{-h/\tau} \right )^2 \\  \frac{1}{h} \left ( 1 - e^{-h/\tau} \right )^2 & \frac{1}{\tau}(1 - e^{-2 h/\tau})  \end{pmatrix}. 
$$
In fact, the Gaussian distribution $\mathcal{N}(xm(h),\Sigma(h))$, which is used at every time step, is the law of the two-dimensional random vector
\begin{equation}
\left ( \frac{1}{h} \int_0^h X_s^{\star,x} { \rm d} s , X_h^{\star,x} \right )
\end{equation}
where we use the notation $X_h^{\star,x}$ for the state of the OU noise variable at time $h$ provided that it started from $x$ at time $0$. 

The MC estimator $\hat D_N^\star$ and a confidence interval for $D^\star$ are built from the sample $\{ U_{\textup{le}}^{\star,(k)}, t_{\textup{le}}^{\star,(k)}\}_{k=1}^N$ by using Equation \eqref{le_estimator} and Appendix~\ref{app:B}.

\section{Numerical results}
\label{sec:num}
This section is devoted to numerical results produced by the algorithms presented in the previous section. We study the sensitivity of the diffusivity $D$ with respect to the strength of the noise $\Gamma$ and the correlation time $\tau$.

\paragraph*{Simulation parameters}
In the results shown below, the differential inclusion \eqref{eq:diff_incl} is integrated with a time step of $h = 10^{-4} \, s$. 
Each Monte Carlo result is produced with $N=10^5$. 
  
\subsection{Comparisons between PDMP and differential inclusion simulations}
\begin{figure}[h!]
\centering
\includegraphics[scale=1]{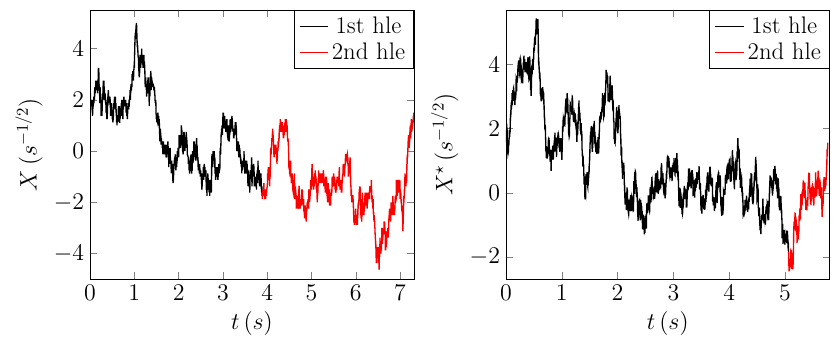}
\includegraphics[scale=1]{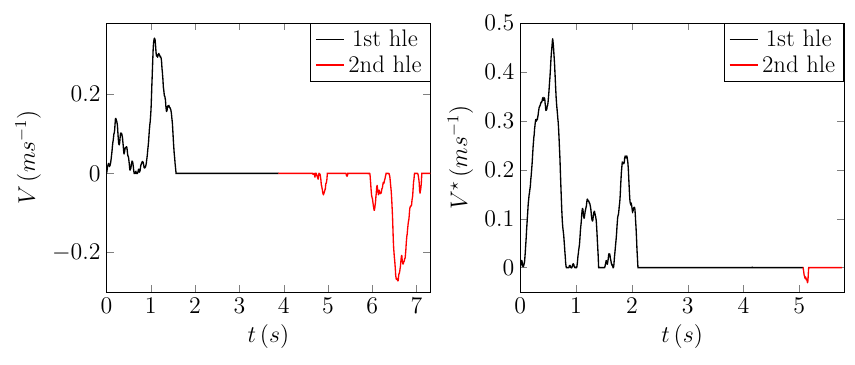}
\includegraphics[scale=1]{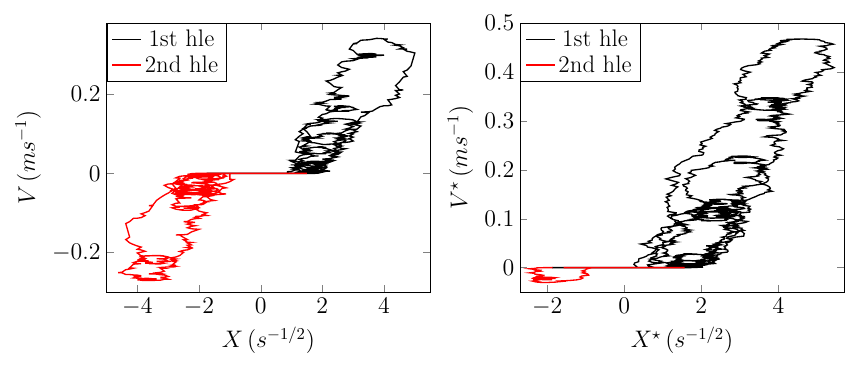}
\caption{
Numerical stochastic dynamics of a long excursion when $\mathfrak{b}(v) = - v/\tau_L + \bar \gamma$. Left column:  single long excursion simulation of the PDMP $(X,V)$ with $\delta = 0.125 \, s^{-1/2}$. Right column: single long excursion simulation of the solution of the differential inclusion. Here $\tau = 0.5 \,s$, $\tau_L = 0.067 \, s$, $\Delta=3.84 \, ms^{-2}$ and $\Gamma = 5 \, m^2s^{-3}$, and $\bar \gamma = 0.342 \, ms^{-2}$.}
\label{fig:2}
\end{figure}

\begin{figure}[h!]
\centering
\includegraphics[scale=1]{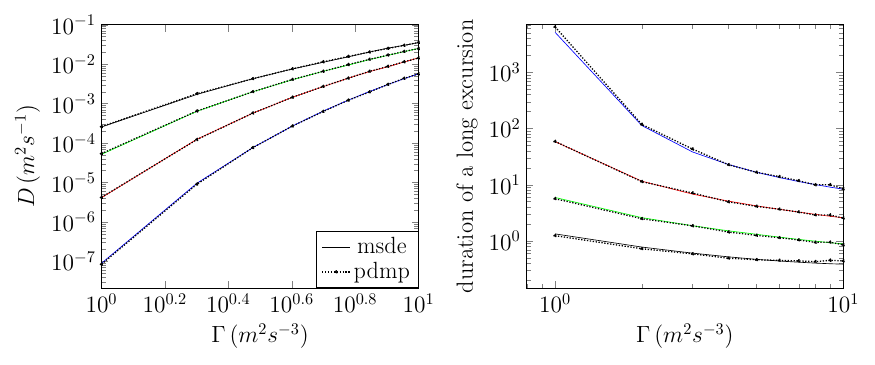}
\caption{Left: Monte Carlo estimation of the diffusivity $D$ as a function of $\Gamma \in [1,10]\, \, m^2s^{-3}$ in loglog scale when $\mathfrak{b}(v) = - v/\tau_L + \bar \gamma$. The dots correspond to numerical simulations of the MC estimator $\hat D_N$ based on the PDMP long excursions when $\delta = 0.125 \, s^{-1/2}$, and the solid lines correspond to the MC estimator $\hat D_N^\star$ based on the long excursions of the limiting differential inclusion as $\delta \to 0$. The four curves from top to bottom correspond to $\tau = 0.125, 0.25, 0.5, 1 \, s$ and we have $\tau_L =0.067 \, s$, $\Delta=3.84 \, ms^{-2}$, and $\bar \gamma = 0.342 \, ms^{-2}$. Right: Monte Carlo estimation of the mean duration of a long excursion as a function of $\Gamma$ in loglog scale. The four curves from bottom to top correspond to $\tau = 0.125, 0.25, 0.5, 1 \, s$. The parameters remain unchanged compared to the left figure. }
\label{fig:3}
\end{figure}

\begin{figure}[h!]
\includegraphics[scale=1]{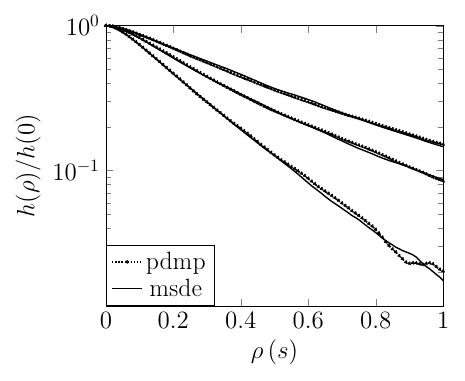}
\caption{$h(\rho)/h(0)$ on a semi-log scale where
$h(\rho) =  \textup{Cov}  (V(t) , V(t+\rho))$ when $\mathfrak{b}(v) = - v/\tau_L + \bar \gamma$.
The solid lines stand for the empirical covariances obtained by PDMP simulations.
The dotted lines stand for the empirical covariances obtained by MSDE simulations.
The three curves from bottom to top correspond to $\tau = 0.25, \, 0.5, \,1 \, s $ (i.e. $\tau'=0.74 , \,1.47, \, 2.95$). 
Here $ \tau_L =0.067 \,s$, $\Delta=3.84 \, ms^{-2}$, $\Gamma = 5 \, m^2s^{-3}$, $\bar \gamma = 0.342 \, ms^{-2}$, and $\delta= 0.125\, s^{-1/2}$ (for the PDMP). 
}
\label{fig:4}
\end{figure}

In Figure \ref{fig:2}, we present a sample of long excursion related to the PDMP $(X,V)$ defined in Section~\ref{sec:pdmp} and the solution of the differential inclusion $(X^\star,V^\star)$ defined by Eq.~\eqref{eq:diff_incl} when $\mathfrak{b}(v) = - v/\tau_L + \bar \gamma$.
Here $\delta=0.125 \, s^{-1/2}$ and  ${\delta'}\simeq 0.073$ which is smaller than one, so we can expect that the distribution of the PDMP solution is close to the one of the limiting differential inclusion.
Indeed, in Figure \ref{fig:2}, the two trajectories have similar behaviors to the naked eye.
In Figure \ref{fig:3}, we superpose the computed diffusivity and the mean duration of long excursions of both $(X,V)$ and $(X^\star,V^\star)$ when $\Gamma \in [1,10] \, m^2s^{-3}$ and $\tau = 0.125,0.25,0.5,1 \, s$. Then, in Figure \ref{fig:4}, we also compute the empirical covariance for each process. 
In agreement with the theory, the statistics of $(X,V)$ are close to those of $(X^\star,V^\star)$ when $\delta$ is small enough (i.e. when ${\delta'}$ is smaller than one).

\paragraph*{Comments.}
The PDMP makes the mathematical framework for the diffusivity very neat. However, one drawback in simulating the PDMP appears when we consider $\tau$ small. Indeed, the jump frequency of the PDMP becomes very high, therefore its dynamics evolves with extremely small time steps. In this context the CPU time becomes significantly important. This is the reason why we extend the notion of long excursion to the limit process in his differential inclusion form.

\subsection{Comparisons between white noise and colored noise regimes}
\begin{figure}[h!]
\includegraphics[scale=1]{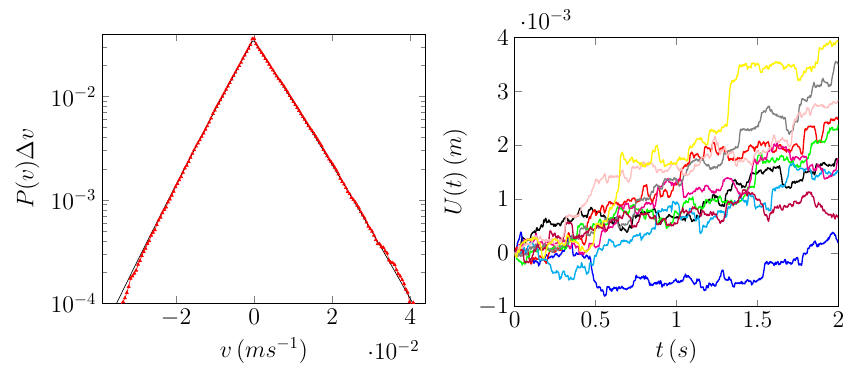}
\caption{Left: comparison between the theoretical stationary probability $P(v) \Delta v$ in solid line and the numerical histogram of the velocity with bin width $\Delta v = 4.8 \, 10^{-4} \, ms^{-1}$ in red triangles for the colored noise driven system when $\mathfrak{b}(v) = - v/\tau_L +\bar \gamma $. 
Right: displacement $U(t)$ versus $t$ for a colored noise driven system. $10$ simulations are plotted on $t \in [0,2] \,s$. Here  $\tau = 10^{-5} \, s$, $\Gamma = 0.16 \, m^2s^{-3} $, $\Delta=3.84 \, ms^{-2}$, $\tau_L = 0.067 \, s$, and $\bar \gamma = 0.342 \, ms^{-2}$.}
\label{fig:proba_and_traj}
\end{figure}

Here we assume that $\mathfrak{b}(v) = - v/\tau_L + \bar \gamma$ and we carry out simulations with the limiting differential inclusion.
As illustrated in Figure \ref{fig:proba_and_traj} (left), the numerically obtained stationary probability for the colored noise with $\tau = 10^{-5} \,s$ agrees with the explicit formula (valid for a white noise) of the theoretical stationary probability \cite{H05} of the velocity
$P(v) = P_0 e^{-v^2/(\Gamma \tau_L) - 2 |v| \Delta/\Gamma + 2 v \bar{\gamma}/\Gamma}$ and $P_0>0$ is a normalizing constant. The white noise regime is indeed expected since {${\tau'}\simeq 9.2 \, 10^{-4}$} is much smaller than one. In addition, some realizations of the dynamics of $U(t)$ are shown for  $\tau = 10^{-5} \, s$ in Figure \ref{fig:proba_and_traj} (right). 
We observe an average positive drift due to the presence of $\bar \gamma$.
This is a good qualitative agreement with Figure 2 of \cite{GMC09}.

\begin{figure}[h!]
\includegraphics[scale=1]{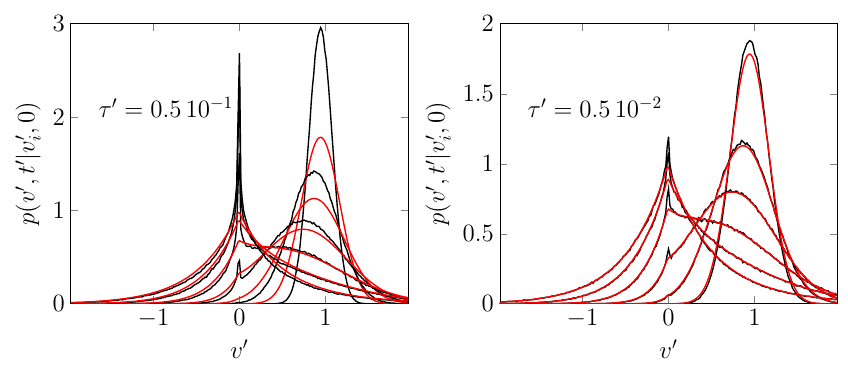}
\includegraphics[scale=1]{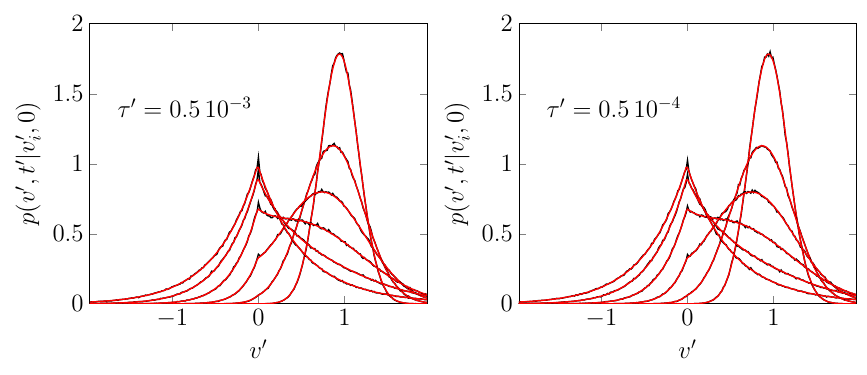}
\caption{Red curves: Probability density function ${v}'\mapsto p({v}', {t}' | {v}_i', 0)$ of the velocity at different times $t'$ for the white noise driven pure dry friction case with initial condition ${v}_i'=1$ at time $0$ in non-dimensional variables
(see Formula (2.10) in \cite{T10}). Black curves: empirical histogram of the velocity for the colored noise driven pure dry friction with the initial condition is $ {v}_i'=1$ and $ {x}_i' \sim \mathcal{N}(0, {\tau'}^{-1})$. The four plots are for four different values of the correlation time $\tau'$.
}
\label{fig:touchette_prediction1}
\end{figure}

\begin{figure}[h!]
\includegraphics[scale=1]{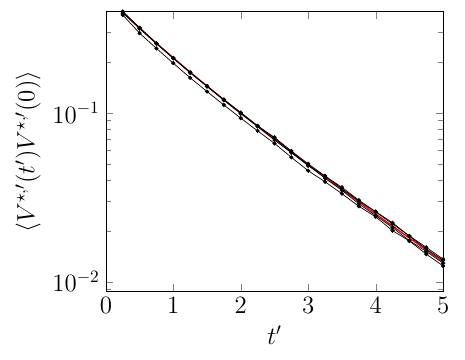}
\caption{Correlation function $\langle V^{\star,\prime}({t'}) V^{\star,\prime}(0) \rangle$ versus ${t'}$ in semilog scale for the pure dry friction case under stationarity. 
The red solid line is the explicit formula (2.13) in \cite{T10} (valid when $ {\tau'} \downarrow 0$). 
There are four curves in black dots from our simulations. 
The curves associated with the colored noise case 
where ${\tau'} = 0.5 \, 10^{-4} $, ${\tau'} = 0.5 \, 10^{-3}$, ${\tau'} = 0.5 \, 10^{-2} $, are almost indistinguishable.  
The remaining curve below the red curve is for ${\tau'} = 0.5 \,  10^{-1}$ and it is also very close to the first three ones.
}
\label{fig:touchette_prediction3}
\end{figure}

Here we consider the pure dry friction $\mathfrak{b}(v) = 0$
and we want to compare our numerical results with the theoretical predictions of \cite{T10}  valid in the white noise regime. We here consider the system in non-dimensional variables.
The numerically obtained histogram, first moment, and correlation function for the velocity $\check{V}^{\star,\prime}({t'})$ are shown for several values of the noise correlation time $ {\tau'}$ (${\tau'}= 0.5 \, 10^{-i} , \: 1 \leq i \leq 4$) in Figure \ref{fig:touchette_prediction1}, Table \ref{tab:touchette_prediction2} and Figure \ref{fig:touchette_prediction3} respectively.
As ${\tau'} \to 0$, all our simulation results capture the predictions of \cite{T10} (see formula (2.10), (2.11) and (2.13) therein).
We can see, however, a significant departure in Figure~\ref{fig:touchette_prediction1} for $ {\tau'}= 0.5 \, 10^{-1} $ which means that the white-noise approximation is no longer valid for such a value of the correlation time to give predictions of the statistics of the velocity.

 
\begin{table}
\begin{tabular}[b]{ c | c c c c c c }
$t'$ & $0.05$ & $0.125$ &  $0.25$ & $0.5$ & $1$ & $2.5$\\ \hline
MC & 0.950 & 0.876  & 0.757 & 0.569 & 0.337 & 0.091\\ 
EF &  0.950  & 0.875 & 0.757 & 0.568 & 0.336 & 0.090
\end{tabular}
 \caption{First moment of $V^{\star,\prime}(t')$ versus $t'$ for the pure dry friction case with initial condition $v_i'=1$. 
The MC line results from our simulations with $\tau' = 0.5 \: 10^{-5}$.
 The EF line is the explicit formula (2.11) in \cite{T10}.}
\label{tab:touchette_prediction2}
 \end{table}
 
\begin{figure}[h!]
\includegraphics[scale=1]{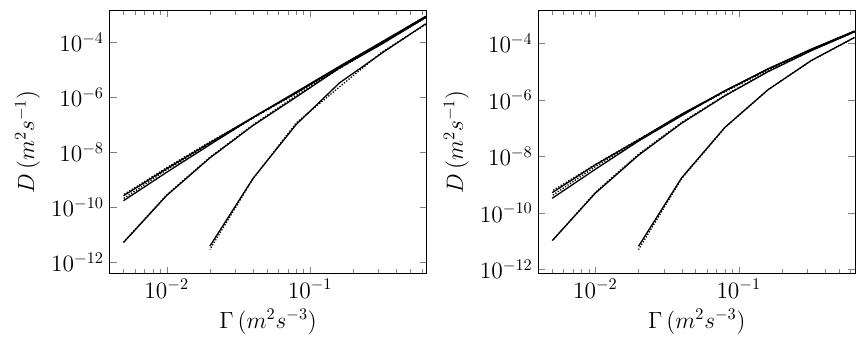}
\caption{Left: case $\mathfrak{b}(v) = 0$. Right: case $\mathfrak{b}(v) = -v/\tau_L + \bar \gamma$ with $\tau_L = 0.067 \, s$ and $\bar \gamma = 0.342 \, ms^{-2}$.
The dots correspond to the numerical simulation of $\hat D_{N}^{t,\star}$ (the MC estimator based on the brute force simulation of the limiting differential inclusion with $t=10 \, s$). The solid lines correspond to the numerical simulation of $\hat D_N^\star$ (the MC estimator based on the long excursion of the limiting differential inclusion). In both cases, the four curves from top to bottom correspond to $\tau = 10^{-i} \, s$ for $i=5, \ldots, 2$.
Here $\Delta=3.84 \, ms^{-2}$.
}
\label{fig:scalinglaw}
\end{figure}

As shown in the left subfigure of Figure \ref{fig:scalinglaw} produced with $\mathfrak{b}(v) = 0$, the diffusivity varies as $\Gamma^3$ when $\tau = 10^{-j}\, s$, $j=4,5$,
i.e. ${\tau}'\simeq 2.9 \, 10^{-j}$, $j=4,5$ (close to white noise). Otherwise when $\tau$ gets larger ($\tau = 10^{-j} \, s$, $j = 2,3$),  the relationship in log log scale between $D$ and $\Gamma$ is not linear and thus there is no scaling law of the form $D \sim \Gamma^\alpha$ with a constant $\alpha$. This means that the white-noise approximation is not valid anymore for $\tau' \simeq 2.9 \, 10^{-j}$, $j=2,3$ to study the diffusivity. The white-noise approximation should be used with caution and even a small correlation time of the driving force can have a strong impact.
As shown in the right subfigure of Figure \ref{fig:scalinglaw}  produced with $\mathfrak{b}(v) = -v/\tau_L + \bar \gamma$, the same comment applies to all the cases for the relationship in log log scale between $D$ and $\Gamma$.

While we recover several theoretical results from Hayakawa \cite{H05}, de Gennes \cite{DG05}, Touchette \cite{T10,T12}, we cannot say the same for the experimental results from \cite{GMC09}.  
In their experimental study 
we have $\mathfrak{b}(v) = -\tau_L^{-1} v + \bar{\gamma}$ 
where $\tau_L \simeq 0.067 \, s$ is the momentum relaxation time and
$\bar{\gamma} \simeq 9.8 \sin \left ({\pi}/{90} \right ) \simeq 0.342 \, ms^{-2}$ is a constant related to gravity and the inclination of the surface on which the system is installed.
The noise in the experiment is assumed to be a white noise and the friction coefficient $\Delta$ is estimated to be $3.84 \, ms^{-2}$.
The experimentally obtained diffusivity scales as $\sim \Gamma^{1.61}$ which is not too far from their simulations predicting a scaling $\sim \Gamma^{1.74}$ where the noise strength $\Gamma$ varies between $5.10^{-3}$ and $5.10^{-1} \, m^2s^{-3}$.
When comparing with our results in the right subfigure of Figure \ref{fig:scalinglaw} we can observe a discrepancy.
We believe that there are two possible (and related) explanations for such a discrepancy.
First the experimental and numerical forces are assumed to be white noises in \cite{GMC09}  and we have exhibited above that the correlation time should be very small to ensure the validity of the white-noise approximation for the study of the diffusivity.
We do not know the correlation time in the experiments, and the correlation time in the numerical simulations in \cite{GMC09} was apparently equal to the integration time step $10^{-3} \, s$, which means that the white-noise approximation does not seem to be valid.
Second we have observed a high sensitivity of the numerical diffusivity to the integration time step itself. In our simulations, we observed that the computation of the diffusivity in Figure~\ref{fig:scalinglaw} appears to be more sensitive to the time step than the computation of the empirical histogram of the velocity in Figure~\ref{fig:touchette_prediction1}. Both Figures~\ref{fig:touchette_prediction1} and \ref{fig:scalinglaw} show results produced with $h = 10^{-4}\, s$. We have observed that the results do not change when we take a smaller $h$. We have observed, however, that the results change when $h$ reaches values of the order of $10^{-3} \, s$. More exactly,  the results of Figure~\ref{fig:touchette_prediction1} do not vary much but  those of Figure~\ref{fig:scalinglaw} vary significantly. It turns out that the acquisition time of the video recording in the experiments and the time step in the numerical simulations in \cite{GMC09} are both of this order of magnitude so it may explain the  discrepancy. 
This observation strengthens the need of accurate simulation methods and makes the use of efficient Monte Carlo methods even more important in the context of expensive numerical simulations.

\section{Conclusions}
\label{sec:conclusion}%
In this paper we have introduced a piecewise deterministic Markov process approach to model the random motion of an object subject to dry friction in presence of colored noise. The latter is represented by a pure jump process that is itself a $\delta$ spatial discretization of an Ornstein-Uhlenbeck noise with correlation time $\tau$. 
In this model we have identified an independent and identically distributed sequence of repeating patterns or excursions. This excursion is the fundamental brick of the dynamics because it encodes all the behavior of the system. We have shown that the variance of the object displacement has linear growth in time. We have obtained a representation formula for the diffusivity (the linear growth rate) as an expectation of a functional of an excursion. As a by product, we have derived a new Monte Carlo estimator for the diffusivity with much better properties that standard Monte Carlo estimators. The method we have developed can be used to calculate quantities similar to diffusivity (e.g. mobility etc) with high accuracy and confidence.

As the PDMP cannot be used for numerical purposes when $\tau$ and $\delta$ are small due to high frequency of jumps, we have extended the notion of excursion to the limit process as $\delta \downarrow 0$. When $\tau \downarrow 0$, all our numerical simulations for the stationary probability density function, the transition probability density function, the first moment, the correlation and the diffusivity are captured by the theoretical predictions of Hayakawa \cite{H05}, de Gennes \cite{DG05}, and Touchette \cite{T10,T12}. We have further investigated these quantities as functions of the correlation time $\tau$ of the noise. 
We have shown that the white-noise approximation gives correct predictions for the distribution of the velocity for small or moderately small values of the correlation time, but the white-noise approximation requires very small values of the correlation time to give correct predictions for the diffusivity. 

\section*{Acknowledgements}
Laurent Mertz thanks NSFC grant 12271364.
The authors would like to thank Carl Xu for useful discussions.

\appendix

\section{The driving jump process}
\label{app:genX}%
The random dynamics of $X$  starting from a state $X(0)=\xi_0$ is as follows.\\
{\bf 1)} Generate a random time $\tau_1$ with an exponential distribution with parameter $\Lambda $. Set $X(t)=\xi_0$ for $ t \in [0,\tau_1)$. \\
{\bf 2)} If $|\xi_0|<x_N$, then with probability $\alpha_{\xi_0}$,  set $\xi_1=\xi_0+\delta$ and with probability $1-\alpha_{\xi_0}$, set $\xi_1=\xi_0-\delta$.\\
If $\xi_0=x_N$, then  set $\xi_1=x_{N-1}$.\\
If $\xi_0=x_{-N}$, then  set $\xi_1=x_{-N+1}$.\\
{\bf 3)} Generate a random time $\tau_2$ with an exponential distribution with parameter $\Lambda$. Set $X(t)=\xi_1$ for $ t \in [\tau_1,\tau_1+\tau_2)$. \\
{\bf 4)} Iterate.  $X$ is piecewise constant, takes values in $S^\delta$, and has random jumps at times $\sum_{i=1}^j \tau_i, j \geq 1$.

\section{Description of the PDMP}
\label{app:PDMP}
We give details on the definition of the PDMP modeling dry friction.

The process $X$ defined in Section~\ref{subsec:markovjump} 
is a jump Markov process with the generator 
$Q^\delta f(x) = 2\tau^{-2} \delta^{-2} \left ( \alpha_x f(x+\delta) - f(x) + (1-\alpha_x) f(x-\delta) \right )$ where $\alpha_x = \frac{1}{2} \big(1 -\frac{ \tau \delta x}{2}\big)$  if $|x| < x_N$,
$0$ if $x=x_N$, $1$ if $x =x_{-N}$.
Here we assume $\tau \delta L_X^\delta <2$ to guarantee that $\forall x \in S^\delta, \alpha_x \in [0,1]$.

We introduce
\begin{equation}
B(x,y,v) =
\begin{cases}
\Delta + \mathfrak{b}(v) + \sqrt{\Gamma}x, & \: \mbox{ if } \: y = -1,\\ 
0, & \: \mbox{ if } \: y = 0,\\ 
-\Delta + \mathfrak{b}(v) + \sqrt{\Gamma}x, & \: \mbox{ if } \: y = 1.
\end{cases}
\end{equation}

We define the state space 
\begin{equation}
E = \bigcup_{(x,y)\in 
\mathbb{S}^\delta 
} E_{x,y}, \quad \quad E_{x,y}= \{(x,y)\}\times H_{x,y},
\end{equation}
where 
$\mathbb{S}^\delta = \{ x_{-N},\ldots ,x_{k_--1}\} \times \{-1,1\} \cup \{ x_{k_-},\ldots ,x_{k_+}\} \times \{-1,0,1\} \cup \{ x_{k_++1},\ldots ,x_{N}\} \times \{-1,1\}$,
$H_{x,y} = (-\infty,0)$ if $(x,y) \in
 \{ x_{k-} ,\ldots, x_N \}$ 
$\times \{-1\}$,
$H_{x,y}$ $=(0,+\infty)$ if $(x,y) \in 
\{ x_{-N},\ldots, x_{k_+} \}$
$\times \{1\}$,
and $H_{x,y}=\RR$ otherwise.

We can formulate the dynamics of $Z$ starting from a state $z_0 = (x,y,z) \in E$ as follows.

{\bf 1)} Generate a random time $T_1 = \min \left ( \tau_1, T^\star(z_0) \right )$
where $\tau_1$ is a random time with an exponential distribution with parameter $\Lambda = 2 \tau^{-2} \delta^{-2}$, $T^\star(z_0) = \inf \{ t \geq 0, \: \phi_{x,y}(t,v) = 0 \}$ (with the convention $\inf \emptyset=+\infty$)
and $\phi_{x,y}(t,v)$ is the flow solution of 
\begin{equation}
\begin{cases}
\partial_t \phi_{x,y}(t,v) = B(x,y,\phi_{x,y}(t,v)) , \: t >0, \\
\phi_{x,y}(0,v) = v.
\end{cases}
\end{equation}
Then define $v_1 = \phi_{x_0,y_0}(T_1,v_0)$ and generate a random state $z_1 = (x_1,y_1,v_1)$ from $(x_0,y_0,v_1)$ using 
the probability transition matrix ${\mathcal Q}_{v_1}(x_1,v_1;x_0,v_0)$ (note that the velocity $V$ does not jump during this transition):
\begin{subequations}
\label{eq:defcalQ}
    \begin{align}
&\nonumber
\forall y \in \{-1,1\}, \: 
\forall x \in \{ x_{-N}, \ldots, x_{k_--1} \},\\
&
\quad {\mathcal Q}_0 \big(  x,-1 ; x,y ) =1,\\
&
\nonumber
\forall y \in \{-1,1\}, \: 
\forall x \in \{ x_{k_++1}, \ldots, x_N \},\\  
&
\quad  {\mathcal Q}_0 \big(  x,1 ; x,y ) =1,\\
&
\nonumber
\forall y \in \{-1,1\}, \: 
\forall x \in \{ x_{k_-}, \ldots, x_{k_+} \},\\ 
& 
\quad {\mathcal Q}_0 \big(  x,0 ; x,y ) =1,\\
\nonumber
&\forall x \in \{ x_{k_-+1}, \ldots, x_{k_+-1} \}, \\  
& 
\quad {\mathcal Q}_0 \big(  x+\delta,0 ; x,0 ) = \alpha_x,\\
\nonumber
& \forall x \in \{ x_{k_-+1}, \ldots, x_{k_+-1} \}, \\ 
& 
\quad {\mathcal Q}_0 \big(  x-\delta,0 ; x,0 \big ) = 1-\alpha_x,\\
& {\mathcal Q}_0 \big( x_{k_-+1},0 ; x_{k_-},0 ) =\alpha_{x_{k-}},\\
& {\mathcal Q}_0 \big( x_{k_+-1},0 ; x_{k_+},0 ) =1-\alpha_{x_{k_+}},\\
& {\mathcal Q}_0 \big( x_{k_--1},-1 ; x_{k_-},0 ) =1-\alpha_{x_{k_-}} , \\
&
{\mathcal Q}_0 \big( x_{k_++1},1 ; x_{k_+},0 ) =\alpha_{x_{k_+}},    \\
\nonumber
&
\forall (x,y,v) \in E,   \: v \neq 0,\\ 
&
\quad  {\mathcal Q}_v \big( x+\delta,y; x,y ) =\alpha_x,\\
&
\nonumber
\forall (x,y,v) \in E,  \:v \neq 0, \\
&
\quad  {\mathcal Q}_v \big( x-\delta,y ; x,y ) =1-\alpha_x.
\end{align}
\end{subequations}

The trajectory of $Z$ for $t \in [0,T_1]$
 is given by 
\begin{equation}
 Z_t = 
 \begin{cases}
(x_0,y_0,\phi_{x_0,y_0}(t,v_0)), & \mbox{ if } \: 0 \leq t < T_1\\ 
(x_1,y_1,v_1) , & \mbox{ if } \: t = T_1.
 \end{cases}
\end{equation}
When $\mathfrak{b}(v) = - v/\tau_L+\bar \gamma$ with $\tau_L \in (0,\infty), \bar \gamma \in \mathbb{R}$ 
and due to the structure of $B$, explicit formula for $\phi_{x,y}(t,v)$ and $T^\star(z)$ are available. 
Straightforward calculations give 
$$
\begin{cases}
& \phi_{x,y}(t,v) = |y| \left ( e^{-t/\tau_L}(v-c(x,y)) + c(x,y) \right ),\\
& c(x,y) = \tau_L \big( \bar \gamma + \sqrt{\Gamma}x - y\Delta \big),
\end{cases}
$$ 
and, using the notation $\Xi = \{ (x,y,v) \in E, v>0 \mbox{ and } x < x_{k_+} \: \mbox{ or } \: v<0 \mbox{ and } x > x_{k_-} \}$, 
$$
T^\star(z) =  
\begin{cases}
& \tau_L \log \big(1 - \frac{v}{c(x,y)} \big), \: \mbox{ if } (x,y,v) \in \Xi , \\
& \infty, \: \mbox{ otherwise}.
\end{cases}
$$ 
Furthermore the corresponding displacement on $[0,T_1]$ is $U(T_1) = \int_0^{T_1} \phi_{x_0,y_0}(t,v_0) \textup{d} t = |y_0|(c(x_0,y_0)T_1 + \tau_L (v_0-c(x_0,y_0))(1-e^{-T_1/\tau_L}))$.

{\bf 2)} We can now define $Z$ after $T_1$. Starting from $Z_{T_1}=z_1$, we generate the next jump time $ T_2 = T_1+ \min \left ( \tau_2, T^\star(z_1) \right )$ where $\tau_2$ is a random time with an exponential distribution with parameter $\Lambda$.  Define $v_2 = \phi_{x_1,y_1}(T_2-T_1,v_1)$ and the post-jump location $z_2=(x_2,y_2,v_2)$ from $(x_1,y_1,v_2)$ using the probability transition matrix ${\mathcal Q}$. The trajectory of $Z$ for $t \in [T_1, T_2]$
 is given by 
\begin{equation}
 Z_t = 
 \begin{cases}
(x_1,y_1,\phi_{x_1,y_1}(t,v_1)), & \mbox{ if } \: T_1 \leq t < T_2 , \\ 
(x_2,y_2,v_2) , & \mbox{ if } \: t = T_2.
 \end{cases}
 \end{equation}
When $\mathfrak{b}(v) = - v/\tau_L+\bar \gamma$ with $\tau_L \in (0,\infty), \bar \gamma \in \mathbb{R}$, the increment of displacement on $[T_1,T_2]$ is $U(T_2) - U(T_1) = |y_1|(c(x_1,y_1)(T_2-T_1) + \tau_L (v_1-c(x_1,y_1))(1-e^{-(T_2-T_1)/\tau_L}))$.

{\bf  3)} Iterate. $Z$ is piecewise deterministic and has random jumps at times $T_j, j \geq 1$.

\section{A technical proof}
\label{app:A}
In order to prove that $R_t$ converges in probability to zero as $t\to +\infty$, we proceed as follows.
We can expand
$$
R_t = \frac{1}{\sqrt{j_t}} \sum_{j=j_t}^{J_t-1} {\mathcal X}_{j} + \tilde{{\mathcal X}}_t , \quad \mbox{ with } \tilde{{\mathcal X}}_t =  \frac{1}{\sqrt{j_t}}  \int_{{t}_{j_t}}^t V(s) ds  .
$$
The variable $\tilde{{\mathcal X}}_t $ goes to zero in probability as $t \to +\infty$ since $\EE_{s_+}[ |\tilde{{\mathcal X}}_t| ] \leq j_t^{-1/2} \EE_{s_+}[ \int_0^{t_1} |V(s)|ds] =O(t^{-1/2})$.\\
By introducing $Y_j =\sum_{j'=j \lfloor t^{1/4}\rfloor}^{(j+1) \lfloor t^{1/4}\rfloor-1} {\mathcal X}_{j_t+j'} $:
$$
\Big| \frac{1}{\sqrt{j_t}} \sum_{j=j_t}^{J_t-1} {\mathcal X}_{j} \Big| \leq 
\frac{1}{\sqrt{j_t}} \sum_{j=-N_t}^{N_t } |Y_{j}|  + O(t^{-1/4}) ,
$$
with $N_t = |J_t -j_t|/\lfloor t^{1/4}\rfloor$. We have $N_t \leq \tilde{N}_t:= t^{1/4+1/16}$ with probability that goes to one as $t\to+\infty$ (because $J_t-j_t = O(t^{1/2})$), therefore, for any $\eps>0$, for $t$ large enough,
\begin{align*}
\PP\Big( \Big| \frac{1}{\sqrt{j_t}} \sum_{j=j_t}^{J_t-1} {\mathcal X}_{j} \Big| \geq \eps\Big) \leq  & \PP ( N_t \geq \tilde{N}_t )\\
& + \PP \Big( \frac{1}{\sqrt{j_t}} \sum_{j=-\tilde{N}_t}^{\tilde{N}_t } |Y_{j}| \geq \eps/2\Big)  .
\end{align*}
The variables $Y_j$ are zero-mean, independent and identically distributed, with $\EE_{s_+}[|Y_1|]\leq \EE_{s_+}[Y_1^2]^{1/2}=  t^{1/8} \EE_{s_+}[{\mathcal X}_1^2]^{1/2}$.
We then get by Markov inequality that
\begin{align*}
\PP\Big( \Big| \frac{1}{\sqrt{j_t}} \sum_{j=j_t}^{J_t-1} {\mathcal X}_{j} \Big| \geq \eps\Big) & \leq 
\PP ( N_t \geq \tilde{N}_t ) + \frac{\frac{2}{\sqrt{j_t}} \sum \limits_{j=-\tilde{N}_t}^{\tilde{N}_t } \EE_{s_+}[ |Y_{j}| ]}{\eps}\\
& \leq \PP ( N_t \geq \tilde{N}_t )  + \frac{C t^{-1/16}}{\eps},
\end{align*}
which shows the desired result:
$$
\PP(|R_t |\geq \eps ) \stackrel{t \to +\infty}{\longrightarrow} 0 .
$$

\section{Asymptotic confidence intervals}
\label{app:B}
In this appendix we show how to build a confidence interval for $D$ defined by (\ref{eq:defD}) from the  sample $\{ (U^{(k)}_{\textup{le}},t^{(k)}_{\textup{le}}) \}_{k=1}^N$.
We remark that 
$$
D = \Psi( \EE_{s_+}[{\itbf X}]), 
$$
with ${\itbf X}=(X_j)_{j=1}^3$, $X_1=U_{\textup{le}} = U(t_1)$, $X_2= {U_{\textup{le}}}^2 = U(t_1)^2$, 
$X_3=t_{\textup{le}} = t_1$, $\Psi({\itbf x})=\frac{x_2-x_1^2}{x_3}$. 
We define 
$$
\hat{\itbf S}_N = \frac{1}{N}\sum_{k=1}^N {\itbf X}^{(k)}  ,
$$
with ${\itbf X}^{(k)}=(X^{(k)}_j)_{j=1}^3$,
$X^{(k)}_1=U^{(k)}_{\textup{le}} $, $X^{(k)}_2= {U^{(k)}_{\textup{le}}}^2$, 
$X^{(k)}_3=t^{(k)}_{\textup{le}}$.
We have $\hat{D}_N=\Psi(\hat{\itbf S}_N)$.
Since the ${\itbf X}^{(k)}$, $k=1,\ldots,N$, are independent and identically distributed, 
we can apply the delta method \cite[p.79]{wasserman} and we get that
the estimator $\hat{D}_N$ converges in distribution:
$$
\sqrt{N}  \big( \hat{D}_N-D\big)
 \stackrel{N \to +\infty}{\longrightarrow} {\cal N}(0, \sigma^2), 
 $$
with $\sigma^2 = \nabla\Psi({\itbf S})^T {\bf C} \nabla \Psi({\itbf S})$,
${\itbf S} = \EE_{s_+}[{\itbf X}]$, ${\bf C}=(C_{jl})_{j,l=1}^3$, $C_{jl} = \EE_{s_+}[ X_j X_l]-\EE_{s_+}[ X_j ]\EE_{s_+}[X_l]$.
Here ${\cal N}(0,\sigma^2)$ stands for the normal distribution with mean $0$ and variance $\sigma^2$.
Denoting
$$
\hat{C}_{N,jl} = \frac{1}{N}\sum_{k=1}^N X_j^{(k)}X_l^{(k)} - \hat{S}_{N,j}\hat{S}_{N,l} ,
$$
the estimator 
$$
\hat{\sigma}_N^2 = \nabla \Psi (\hat{\itbf S}_N)^T \hat{\bf C}_N \nabla  \Psi (\hat{\itbf S}_N) 
$$
converges to $\sigma^2$ in probability. By Slutsky's theorem we get that
$$
\sqrt{N} \hat{\sigma}_N^{-1} \big( \hat{D}_N-D\big)
 \stackrel{N \to +\infty}{\longrightarrow} {\cal N}(0, 1), 
 $$
in distribution. This gives that the interval
$$
\big( \hat{a}_N,\hat{b}_N \big) = 
\Big( \hat{D}_N -q_{1-\alpha/2} \frac{\hat{\sigma}_N}{\sqrt{N}}, \hat{D}_N + q_{1-\alpha/2} \frac{\hat{\sigma}_N}{\sqrt{N}}\Big),
$$
with $q_{1-\alpha/2}$ the $(1-\alpha/2)$-quantile of the distribution ${\cal N}(0,1)$,
is a confidence interval of asymptotic level $1-\alpha$:
$$
\lim_{N\to +\infty}
\PP\Big( D \in \big(\hat{a}_N,\hat{b}_N\big)\Big)  =1-\alpha .
$$

\section{Algorithms related to the PDMP}
\label{app:C}
In this appendix we give the detail of the functions  interjump$(X,Y,V)$, displacement$(X,Y,V)$, \textup{flow}$(X,Y,V,\delta T)$ and jump$(X,Y; V)$ for PDMP simulation. The formulas are valid when $\mathfrak{b}(v) = - v/\tau_L+\bar \gamma$ with $\tau_L \in (0,\infty), \bar \gamma \in \mathbb{R}$. 

\begin{algorithm}[H]
\SetAlgoLined
\KwResult{$\delta T=\textup{interjump}(X,Y,V)$}
$u$ = uniform() \;
$\delta T= \min \left ( - \frac{\log(u)}{\Lambda}, T^\star(X,Y,V) \right ).$ 
\caption{Simulation of an interjump time from $(X,Y,V)$}
\end{algorithm}
\begin{algorithm}[H]
\SetAlgoLined
\KwResult{$\delta U=\textup{displacement}(X,Y,V,T,T+\delta T)$}
$\delta U =  |Y|(c(X,Y)\delta T + \tau_L (V-c(X,Y))(1-e^{- \delta T/\tau_L}))$\;
\caption{Formula for the increment of displacement from $(X,Y,V)$ on the time interval $[T,T+\delta T]$}
\end{algorithm}
\begin{algorithm}[H]
\SetAlgoLined
\KwResult{$\hat V=\textup{flow}(X,Y,V,\delta T)$}
$c(X,Y) = \tau_L \left ( \bar \gamma + \sqrt{\Gamma}X - Y\Delta \right )$\;
$\hat V =  |Y|(e^{-\delta T/\tau_L}(v-c(X,Y)) + c(X,Y))$\;
\caption{Formula for the flow from $(X,Y,V)$ on the time interval $[T,T+\delta T]$}
\end{algorithm}

\begin{algorithm}[H]
\SetAlgoLined
\KwResult{$(X',Y',V)=\textup{jump}(X,Y;V)$}
$\alpha= \frac{1}{2} \left ( 1- \frac{ \tau \delta X}{2} \right ) \mathbf{1}_{\{ |X| < L_X^\delta \}}
+ (1-X^{-1} \max(X,0) )  \mathbf{1}_{\{ |X| = L_X^\delta \}}$\;
$A = (|Y|=1) \mbox{ and } (V=0)$\;
$B = (Y=0) \mbox{ and } (V=0) \mbox{ and } ( (X=x_{k_-}) \mbox{ or } (X=x_{k_+}))$\;
\eIf{A}{$Y'=-\mathbf{1}_{ \{X \leq x_{k_--1}\}}\mathbf{1}_{ \{ Y=1 \} } + \mathbf{1}_{ \{X \geq x_{k_++1}\}} \mathbf{1}_{\{ Y=-1 \}}$}{
$u$ = uniform()\; 
$X'=(X+\delta) \mathbf{1}_{\{ u \leq \alpha \}} +(X-\delta) \mathbf{1}_{ \{ u > \alpha \}}$;\\
\If {B}{
$Y'=\mathbf{1}_{ \{ X=x_{k_+} \} } \mathbf{1}_{\{ u \leq \alpha \}} 
- \mathbf{1}_{ \{ X=x_{k_-} \} } \mathbf{1}_{\{ u > \alpha \}};$}} 
\caption{Simulation of a jump from $(X,Y,V)$}
\end{algorithm}
 
\end{document}